# Describing the Persistence Landscape for Introducing Microbes into Complex Communities

Jason E. McDermott[1,2], William C. Nelson[1], Amy E. Zimmerman[1], Winston Anthony[1], Devin Coleman-Derr[3,4], Joshua Elmore[1], Tara Nitka[1], Ryan S. McClure[1], Pubudu P. Handakumbura[5], Adam Guss[6], Travis J. Wheeler[7], Robert G. Egbert[1]

1 Biological Sciences Division, Pacific Northwest National Laboratory, Richland WA
2 Department of Molecular Microbiology and Immunology, Oregon Health & Science University, Portland OR
3 Plant Gene Expression Center, USDA-ARS, Albany CA
4 Plant and Microbial Biology Department, University of California, Berkeley, CA
5 Environmental Molecular Sciences Division, Pacific Northwest National Laboratory, Richland WA
6 Biosciences Division, Oak Ridge National Laboratory, Oak Ridge, TN
7 College of Pharmacy, University of Arizona, Tucson AZ

## Abstract

The introduction of non-native organisms into complex microbiome communities holds enormous potential to benefit society. However, microbiome engineering faces several challenges including successful establishment of the organism into the community, its persistence in the microbiome to serve a specified purpose, and constraint of the organism and its activity to the intended environment. A theoretical framework is needed to represent the complex interactions that drive these dynamics. Building on the concept of the community functional landscape, we define the 'persistence landscape' as the metabolic, genetic, and broader functional composition and ecological context of the target microbiome that can be used to predict the environmental fitness of an introduced organism. Here, we discuss critical aspects of persistence landscapes that impact interactions between an introduced organism and the target microbiome, including the community's genetic and metabolic complementation potential, cellular defense strategies, spatial and temporal dynamics, and the introduced organism's ability to compete for resources to survive. Finally, we highlight important knowledge gaps in the fields of microbial ecology and microbiome engineering that limit characterization and engineering of persistence landscapes. As a model for understanding microbiome structure and interaction in the context of microbiome engineering, the persistence landscape model should enable development of novel containment approaches while improving controlled colonization of a complex microbiome community to address pressing challenges in human health, agronomy, and biomanufacturing.

## Introduction

Naturally occurring microbes have demonstrated great potential in laboratory studies to promote functions beneficial to host microbiomes such as the human gut or plant rhizosphere. However, their performance in complex environments has been variable due to the difficulties and uncertainties around achieving stable, long term persistence of the organisms in the target community (Mawarda, Lakke et al. 2022). Incorporating new functions by genome engineering into organisms for agronomic, medical, or environmental applications holds great promise for real-world applications (National Academies of Sciences 2019, Afridi, Fakhar et al. 2022, Jansson, McClure and Egbert 2023), but often results in reduced fitness, especially in natural environments. For example, in a field study of grain sorghum inoculated with a synthetic microbial community, the synthetic community did not persist nine weeks after planting despite eliciting increased plant biomass and grain yields (Fonseca-Garcia, Pettinga et al. 2024). However, the incorporation of a genetically engineered microorganism into a soil microbiome, using a genetically tagged *Pseudomonas fluorescens* SBW25 showed that cells persisted up to 250 days after sowing (Jaderlund, Hellman et al. 2008). These contrasting outcomes highlight the challenges with attaining stable, long-term persistence of an introduced organism in natural microbiomes.

In other cases, organisms with strong colonization potential such as pathogens could displace members of the naturally occurring community, thus causing disease, as can happen with *Clostridium difficile* in the human gut (Spigaglia 2024). Better understanding of how pathogenic organisms establish colonization and persist in natural microbiomes would allow design of probiotic strategies for combatting infectious disease, and would provide a basis for biosecurity in agricultural applications like food crops and livestock (Huber, Andraud et al. 2022, Etherton, Choudhury et al. 2024). However, the understanding of how pathogens interact with host microbiomes is poorly understood, especially outside of human pathobiology

The increasing pace of major breakthroughs in synthetic biology (National Academies of Sciences 2019) presents unprecedented opportunities to engineer beneficial functions into microbes deployed in the environment. For example, advances in high-throughput sequencing, DNA synthesis, and methylome profiling technologies have enabled construction of synthetic metabolic pathways, protein engineering, and large combinatorial DNA variant libraries to remodel microbes for novel or enhanced native functions (Zuniga, Fuente et al. 2018, Haskett, Tkacz and Poole 2021, Jansson, McClure and Egbert 2023). DNA editing technologies, such as recombination-mediated genetic engineering, site-specific DNA integration with phage recombinases, inducible natural transformation, and CRISPR/Cas-based genome edits have enabled scientists to implement these engineering solutions in platform organisms and environmental isolates (Egbert, Rishi et al. 2019, Vo, Ronda et al. 2021, Elmore, Dexter et al. 2023). However, an intellectual framework is needed to guide the technical and ethical application of synthetic biology to microbial community engineering.

Successful deployment of both wild and genetically modified microbes in environments requires the consideration of how the microbe interacts with both the desired target community as well as other non-target communities (Wu, Wu et al. 2015, Isabella, Ha et al. 2018, Reardon 2018, Pivot Bio 2019, Qiu, Egidi et al. 2019, Temme, Tamsir et al. 2019, Leventhal, Sokolovska et al. 2020). This requires addressing two central concepts: **colonization** and **control (Table 1)**. Colonization refers to the capacity of the organism to establish in the community in order to carry out its introduced functionality. Control refers to constraining the persistence of the organism to the intended environment and community, preventing proliferation beyond those bounds.

An organism introduced to a target community must be able to successfully acquire carbon, energy, and nutrients needed to survive and persist, while still performing the desired function. It must be competitive with the other community members for access to these resources to prevent being outcompeted. The organism must also be compatible with the members of the target community, avoiding being eliminated by bacterial defense or predation. Understanding these interactions fully would allow engineering mechanisms to both promote persistence of the introduced organism in the target community, while also preventing the organism from growing in other non-target communities.

The control of released microbes is often approached through the design of genetic circuits to contain engineered functions. A common example is genetic modifications that introduce lethal toxins or nucleases (e.g., CRISPR-Cas complexes) upon environmental triggers (Chan, Lee et al. 2016, Rottinghaus, Ferreiro et al. 2022), also referred to as kill switches. More generally, the metabolic and regulatory repertoire of an organism, natively or through gene insertions and deletions, can be leveraged to exert persistence control. One attractive mechanism for engineered control is the removal of a core essential functionality, such that the organism can only survive if the missing function is complemented by some other organism in the target community, but not in non-target communities. This approach is less explored than the use of synthetic genetic circuits and is likely to be more compatible with existing regulatory frameworks (Marken, Maxon and Murray 2024), including in the United States.

The ability of communities to complement functions missing or removed from an introduced organism and support the growth of the organism contributes to the effectiveness of both colonization and control. On one hand, complementation of a deficiency by a member of a target community (e.g., rhizosphere) may allow the organism to successfully colonize and provide beneficial functions (e.g., plant growth promotion). On the other hand, if a deficiency intended for control is complemented by a non-target community (e.g., bulk soil or waterways), the introduced organism may proliferate beyond the target environment with potential for unintended impacts. To describe these complex dynamics, we introduce the concept of a **persistence landscape** – which describes the intersection of the potential of an existing target community to support the establishment and persistence of an introduced organism and that potential of that organism to persist in the community. The persistence landscape, therefore, is tightly linked to mechanisms of control of the introduced organism by engineering or other means. In this perspective we discuss several ways to characterize a persistence landscape, the impact of temporal dynamics and spatial organization on the persistence landscape, and knowledge gaps to address how to enable persistence control of engineered organisms aimed at advancing biotechnology for human health, agriculture, and biomanufacturing.

**Table 1. Glossary of term definitions related to the Persistence Landscape**

| | |
|---|---|
| Establishment | The colonization of a community by an organism |
| Persistence | Colonization in space and time sufficient to perform a specified function |
| Escape | An introduced organism overcoming a control mechanism |
| Introduced organism | A wild or engineered organism presented to a community for colonization |
| Target community | The community intended for controlled colonization |
| Non-target community | A neighboring community that may promote colonization outside the intended environment |

## Defining the persistence landscape

The fitness landscape concept was proposed in the 1930's to help understand the process by which genetic modifications in a single organism result in a novel phenotype that impacts fitness of that organism under a set of defined conditions (Wright 1932). In a classic fitness landscape concept, points on the metaphorical two-dimensional surface represent candidate high-dimensional genotypes of the organism, with similar genotypes being closer together on the surface, and the height of the surface represents the fitness of the organism's phenotype under a particular set of conditions. Represented in this way the metaphor is a powerful way of envisioning a set of high-dimensional relationships that leads to practical insights into non-obvious implications of the model. For example, the shape and ruggedness of the landscape has implications for organismal evolution (Kauffman and Levin 1987) and can be used to identify pairs (or even sets) of genes that affect phenotype, and thus fitness, in non-additive ways, termed epistatic interactions (Diaz-Colunga, Skwara et al. 2023).

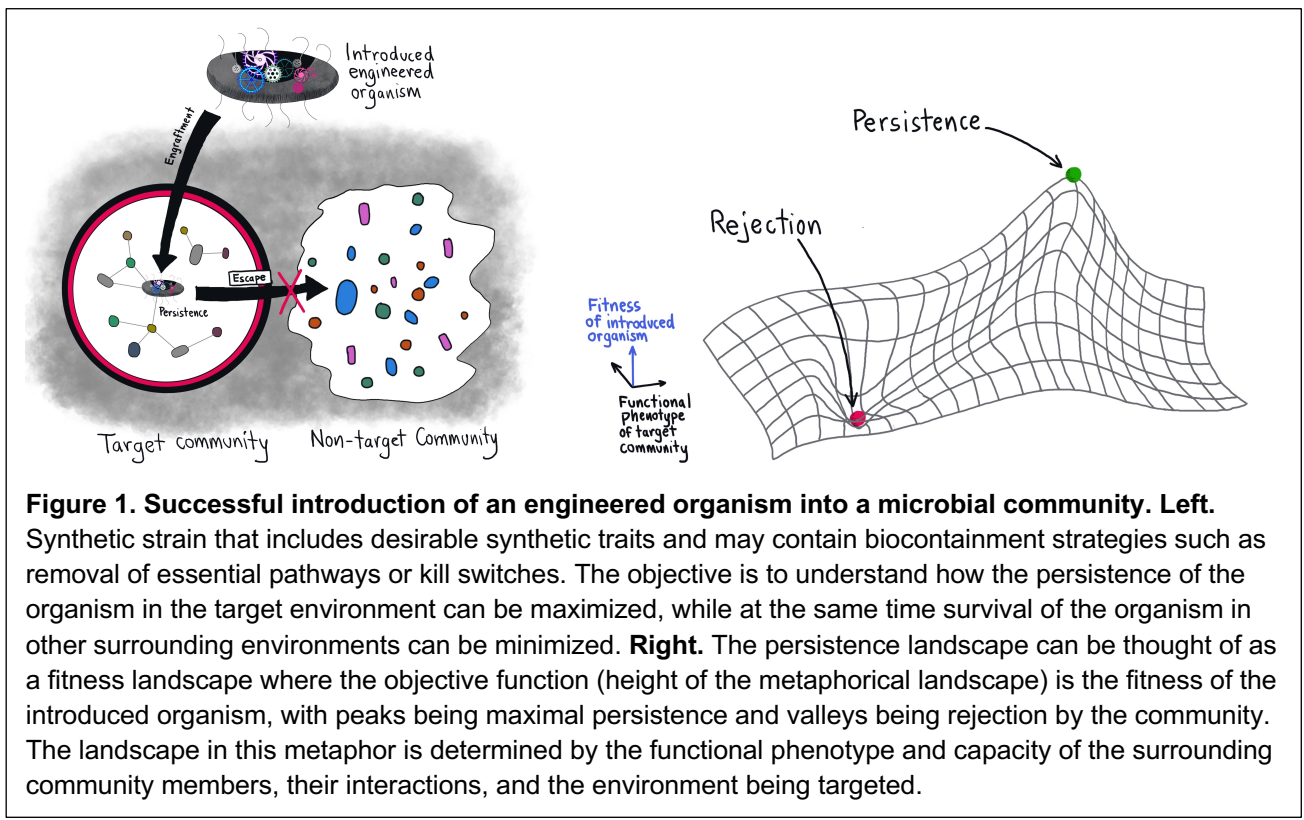

**Figure 1. Successful introduction of an engineered organism into a microbial community. Left.** Synthetic strain that includes desirable synthetic traits and may contain biocontainment strategies such as removal of essential pathways or kill switches. The objective is to understand how the persistence of the organism in the target environment can be maximized, while at the same time survival of the organism in other surrounding environments can be minimized. **Right.** The persistence landscape can be thought of as a fitness landscape where the objective function (height of the metaphorical landscape) is the fitness of the introduced organism, with peaks being maximal persistence and valleys being rejection by the community. The landscape in this metaphor is determined by the functional phenotype and capacity of the surrounding community members, their interactions, and the environment being targeted.

Recently, the notion of the fitness landscape has been extended to microbial communities as community function landscapes (Sanchez, Bajic et al. 2023), which generally describe the effects of species composition and abundance (and associated genetic potential) – the two-dimensional space – on a functional outcome, e.g. the ability to metabolize a specific carbon source (Sanchez-Gorostiaga, Bajic et al. 2019), forming the 3D landscape. Applications of the functional landscape concept to simple engineered communities identified critical species in the community, interactions with importance under particular conditions, and the sensitivity of the community composition to carbon source metabolism (Sanchez-Gorostiaga, Bajic et al. 2019, Sanchez, Bajic et al. 2023).

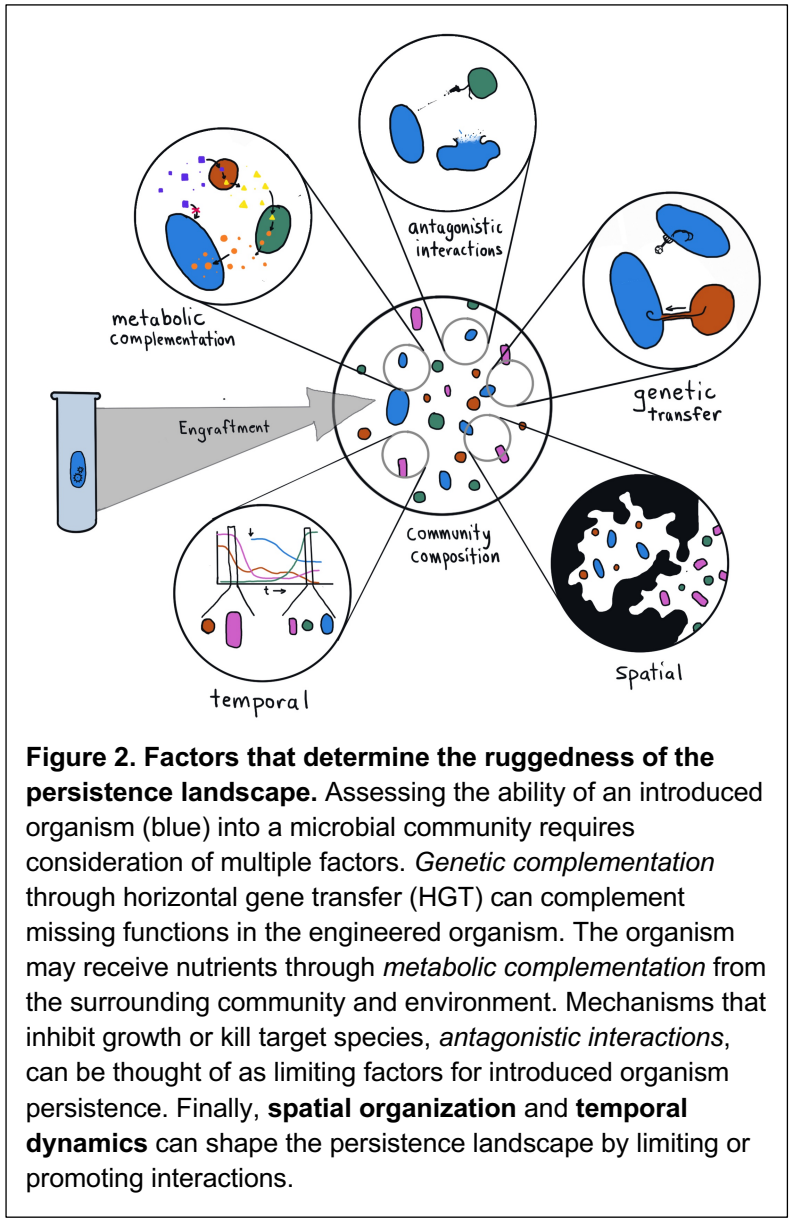

**Figure 2. Factors that determine the ruggedness of the persistence landscape.** Assessing the ability of an introduced organism (blue) into a microbial community requires consideration of multiple factors. *Genetic complementation* through horizontal gene transfer (HGT) can complement missing functions in the engineered organism. The organism may receive nutrients through *metabolic complementation* from the surrounding community and environment. Mechanisms that inhibit growth or kill target species, *antagonistic interactions*, can be thought of as limiting factors for introduced organism persistence. Finally, **spatial organization** and **temporal dynamics** can shape the persistence landscape by limiting or promoting interactions.

We extend the concept of the community function landscape to the **persistence landscape**, where the 'function' being assessed is the ability of an introduced organism to establish and persist in the new environment. In our framing each point on the high dimensional surface of the landscape (Figure 1) is formed by the functions represented in the target microbiome, and the potential for supporting or hindering establishment and persistence of a specific introduced organism as the height of the surface. In the context of bioengineering, the application of the persistence landscape concept would improve identification of the factors and interactions between community members that facilitate microbiome colonization by introduced organisms and inform the design of containment strategies that restrict escape through complementation potential outside the target community.

## Community Composition as a Factor for Persistence

Measuring the composition of a community is the most approachable way to estimate a persistence landscape. Characterization of taxonomic composition for a microbiome is commonly accomplished through amplicon sequencing, which can resolve the identity of bacteria to the species level and fungi to the family level (Hugerth and Andersson 2017). Taxonomic identification by amplicon sequencing provides a low-resolution view of the microbiome, but it can serve as a proxy for the functional potential encoded in related genomes (Martiny, Treseder and Pusch 2013, Martiny, Jones et al. 2015, Douglas, Maffei et al. 2020). However, these methods assume that function does not change over time or space and does not account for the diversity of functions encoded by a species' accessory genes. Therefore, while it can serve as a starting point, this level of information is of limited utility for determining interactions between community members.

Computational models of population dynamics can enhance understanding of interactions between community members and thus help characterize a persistence landscape. Dynamic network models have been used to represent microbial abundance data from temporally resolved 16S rRNA sequences and predict interactions within mouse gut microbiomes (Steinway, Biggs et al. 2015). Likewise, generalized Lotka-Volterra models – which describe ecological predator-prey relationships – have been widely used to model microbiomes at the species abundance level (Stein, Bucci et al. 2013, Gonze, Coyte et al. 2018, Hu, Cai et al. 2024). The Minimal Interspecies Interaction Adjustment (MIIA) method (Lee, Haruta et al. 2019) predicts interactions between community members using a population-based model. Ecological models based on species abundances, but supplemented with genome-scale metabolic models, have been used in the human gut to predict networks of metabolic exchange (Goyal, Wang et al. 2021). These tools provide a starting point for identifying community interactions that may be important for predicting the persistence of an introduced organism.

## Beyond Community Composition

Though community composition may be a tractable experimental target for evaluating the persistence landscape of very simple communities, this information does not reveal the genomic, metabolic, or ecological mechanisms of the interactions observed in the specific environment. In the context of microbiome engineering, such measures are unlikely to suggest strategies for colonization or control. To understand the persistence landscape at a mechanistic level, functional characterization of the introduced organism and target community is necessary. One broad category of interactions that impacts persistence is the complementation of functional deficits.

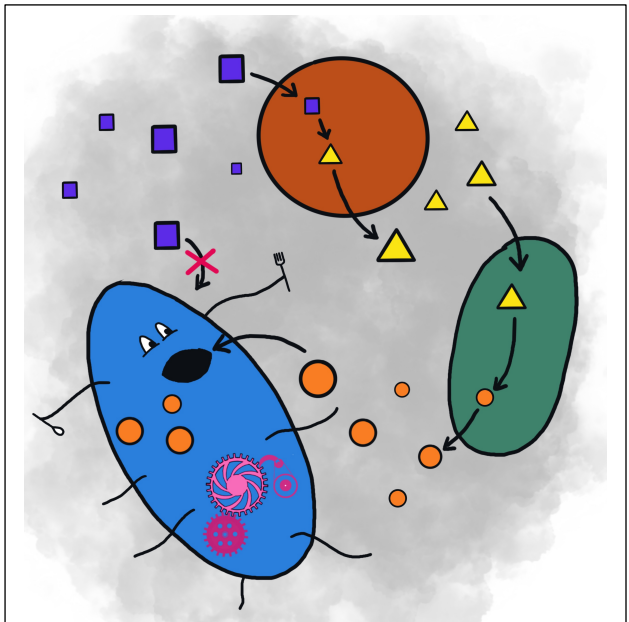

**Figure 3. Metabolic complementation.** Metabolites that can be used by an introduced organism may be made available by the surrounding community members as pathway intermediates, degradation products, or waste products.

Complementation can occur by several mechanisms. **Metabolic complementation** occurs when a metabolite is made available by the surrounding community members, such that a missing function in an introduced organism is complemented through metabolic exchange or syntrophy (Zelezniak, Andrejev et al. 2015). **Antagonistic interactions** describe the interplay between microbial chemical or protein weapons and defenses (Chassaing and Cascales 2018). **Genetic complementation** occurs through horizontal gene transfer (HGT), where genetic material encoding a function missing from an introduced organism is acquired from the community through methods such as conjugation, the uptake of environmental DNA, or through the action of bacteriophage (Tokuda and Shintani 2024). We consider these mechanisms of complementation and interaction as they pertain to persistence in the following sections.

## Metabolic Complementation Potential

A key facet of the persistence landscape is understanding the extent to which functions of an introduced organism essential to survival in its environment can be complemented through metabolic exchange with the target community (to promote persistence) and with non-target communities (that enable escape).

Metabolic exchange (Figure 3) is the assimilation of a compound or by-product by one organism from another organism (Mori, Ponce-de-Leon et al. 2016, Geesink, Ter Horst and Ettema 2024). Multiple exchange mechanisms have been documented, ranging from one organism's waste product being used by another, to metabolic pathways for synthesis of (e.g.) amino acids being partitioned between multiple organisms (Sloan and Moran 2012, Van Leuven, Meister et al. 2014, Miller, Fitzsimonds and Lamont 2019, Garber, Garcia de la Filia Molina et al. 2024). Similarly, polymer degradation can depend on metabolic exchange of breakdown products between specialists and generalists, as in the case of chitin degradation by soil bacteria (McClure, Farris et al. 2022). Metabolic complementation has been well documented in endosymbiotic relationships (Baumann 2005, Van Leuven, Mao et al. 2019, Garber, Garcia de la Filia Molina et al. 2024), with complex interdependent molecular interchange (Garber, Kupper et al. 2021) enabling dramatic reduction in endosymbiont genome size (McCutcheon and Moran 2011, McCutcheon, Garber et al. 2024). Interactions at the scale of complex communities have also been documented across human-associated, soil, and seawater microbiomes (Zelezniak, Andrejev et al. 2015, McCutcheon and Lekberg 2019, Coe, Braakman et al. 2024).

Understanding a community's capacity for metabolic complementation will be critical for determining the persistence landscape and will guide organism engineering strategies. Empirical approaches to evaluate metabolic interactions among community members include generating high-throughput omics data such as metagenomics, metatranscriptomics, and metaproteomics. Statistical inference methods can be applied to these datasets to predict interactions between different species, including metabolic interactions (McClure, Overall et al. 2018, McClure, Farris et al. 2022, Newman, Macovsky et al. 2024, Ruiz-Perez, Gimon et al. 2024). More direct experimental assessment of metabolic exchange can be accomplished using stable isotope probing (SIP) techniques that trace isotopically labeled compounds to different members within a microbial community. For example, metabolism of labile or structural plant derived compounds by specific members of the soil microbiome was measured by incorporation of $^{13}$C-labeled xylose and cellulose (Pepe-Ranney, Campbell et al. 2016). In a host-associated system, $^{13}$C tracer was paired with metabolomics to track metabolites as they were exchanged between community members both *in vitro* and *in vivo* (Henriques, Dhakan et al. 2020). SIP methods have been applied in conjunction with secondary ion mass spectrometry (nanoSIMS) chemical imaging and species identification using fluorescent *in situ* hybridization (FISH) methods, termed 'nanoSIP', to add spatial resolution to isotope tracing (Musat, Musat et al. 2016). These studies provide foundational tools to understand mechanisms of complementation interactions, where a nutrient of interest could be tracked to quantify transfer to an introduced organism. While the scale of SIP experimental approaches is currently limited by expensive reagents and labor-intensive methods, efforts to overcome these challenges (Nuccio, Blazewicz et al. 2022) have the potential to advance tracking metabolite exchanges among individual organisms in complex communities.

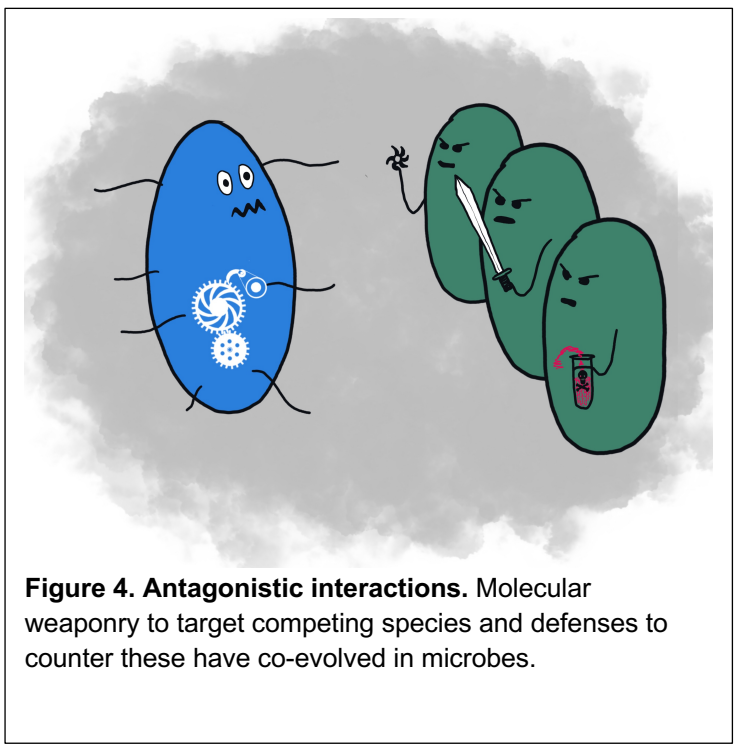

**Figure 4. Antagonistic interactions.** Molecular weaponry to target competing species and defenses to counter these have co-evolved in microbes.

Computational methods developed to enable the prediction of metabolic exchange within microbiomes could be applied to help determine a persistence landscape. Genome-scale metabolic modeling can be used to infer potential metabolic exchange by testing for complementary pathways in reconstructed metabolic networks from combinations of bacteria that are present in the same community, e.g. (Ang, Lakshmanan et al. 2018, Mirzaei and Tefagh 2024). These approaches are limited by the requirement for complete or near complete genome sequences for community members, limited knowledge about transport mechanisms for intermediates, and the complexity of metabolic mechanisms present for microbes in communities. Despite these caveats, a number of algorithms have been developed and applied to microbiomes to conduct flux analysis from genome-scale metabolic models derived from metagenome data to predict metabolic interaction potential, covering interactions between several to hundreds of microbial taxa (Zomorrodi and Maranas 2012, Zelezniak, Andrejev et al. 2015, Joseph, Zafeiropoulos et al. 2024, Lange, Kranert et al. 2024, Mirzaei and Tefagh 2024). Models to predict metabolite exchange are useful but limited. Improved methods that integrate available data from omics or phenotypic measurements are needed to more fully understand these interactions and ground truth potential interactions in empirical observations. Most omics techniques cannot be applied at a fine enough spatial scale to discriminate the contributions from individual organisms, though advances in single-cell omics are closing that gap (Wu, Velickovic and Burnum-Johnson 2024). Tracking metabolite exchanges among multiple organisms at once also remains a challenge and an opportunity where emerging AI-based approaches may be fruitful (Dama, Kim et al. 2023).

## Antagonistic Interactions

Factors beyond metabolic exchange should also be considered as part of the persistence landscape. Many microbial community members encode functions that limit the ability of an introduced organism to persist in the target community.

For example, microorganisms have evolved multiple mechanisms to target other microbes with molecular weapons (Chassaing and Cascales 2018). Offensive strategies include diffusible factors such as bacteriocins – proteins or peptides produced by bacteria that can inhibit or kill other bacteria – or other compounds with bacteriostatic or bactericidal properties, such as small molecules or non-ribosomal peptides (Chassaing and Cascales 2018, Wu, Pang et al. 2022). In contrast to factors that are delivered by secretion and diffusion, some are mediated by contact dependent inhibition (CDI) mechanisms (Ikryannikova, Kurbatov et al. 2020). In general, bacteriocins and CDIs act through receptors and thus have a limited range of targeted species, usually including those that are closely related to the producing organism (Ghequire and De Mot 2015, Chassaing and Cascales 2018). However, non-peptide antibiotics can have a much wider range for killing, requiring specific resistance mechanisms to prevent their action. Studies have also demonstrated that antimicrobial functions can be carried by phage and other mobile genetic elements, straddling genetic and functional complementation (Dragos, Andersen et al. 2021). Thus, defense against these anti-microbial systems are well-known to jump between organisms through HGT (Ferreiro, Crook et al. 2018, Ruiz-Perez, Gimon et al. 2024), with further potential consequences to persistence and containment.

Microbes have also evolved a myriad of systems providing defense against molecular weapons including antibiotics, and phage infection. Defense against the action of antibiotics commonly can occur through the action of efflux pumps, which can move antibiotics, or other harmful molecules, out of the cytoplasm of affected cells (Nikaido and Pages 2012). Enzymes, such as β-lactamases, can neutralize classes of antibiotics, rendering them inactive and protecting the microbe (Laub and Typas 2024). A number of defense systems aimed at foreign DNA from phage or plasmids have also been described. This includes restriction endonucleases, which cleave DNA based on set patterns (Ershova, Rusinov et al. 2015) and CRISPR-Cas systems, which incorporates the patterns of DNA into the host genome, then targets and cleaves DNA with those patterns when they are presented again (Laub and Typas 2024).

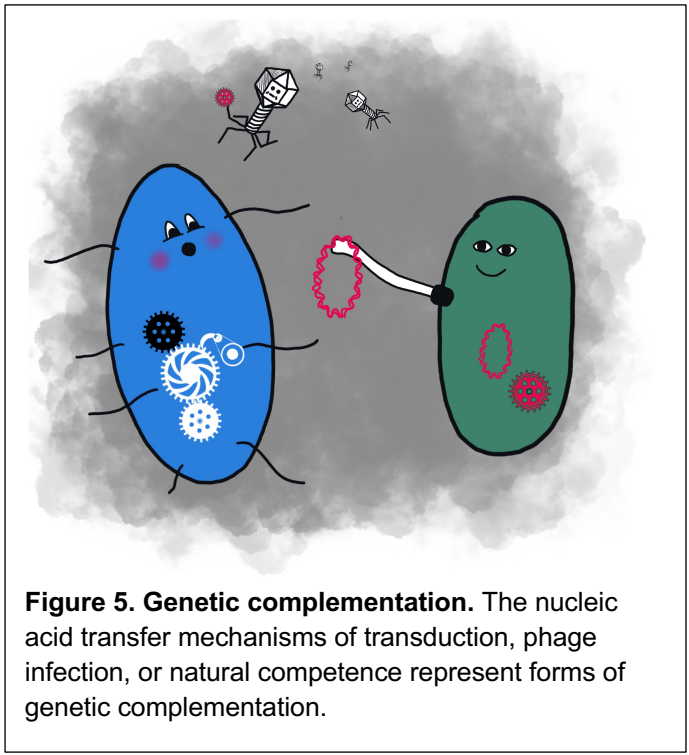

**Figure 5. Genetic complementation.** The nucleic acid transfer mechanisms of transduction, phage infection, or natural competence represent forms of genetic complementation.

Though antagonistic and defensive interactions have high potential to impact a persistence landscape, the limited understanding of the biology and lack of analysis tools in this area represent a major knowledge gap. Computational methods have been developed to identify genes encoding the molecular mechanisms for bacterial warfare and defense, including some bacteriocins (Hammami, Zouhir et al. 2010, Akhter and Miller 2023, Costa, da Silva Moia et al. 2023, Mondal, Sen et al. 2023), but knowledge is limited on the extent of other types of offensive and defensive systems, their ecological and biological target range, and how target identification occurs. Further characterization of how offensive and defensive systems evolve in response to each other and the mechanisms of action are needed. Emerging tools such as protein large language models (Rives, Meier et al. 2021) could be useful to more accurately and sensitively identify these systems and components from genomic data enabling a more thorough characterization of defense complementation and provide insights for manipulating these systems to support persistence.

## Genetic Complementation Potential

HGT describes the process of genetic material being transferred between two organisms (Figure 5). In bacteria, HGT can occur through various mechanisms including direct uptake of DNA from the environment (competence), transfer of DNA directly between two bacteria (conjugation), or transfer via infection by phage (transduction) (Tokuda and Shintani 2024). These processes can shape the persistence landscape in significant ways by providing a mechanism for organisms to acquire functions from other members in the community. We consider the case here where HGT from a native community member to the introduced organism results in the complementation of an absent function in the recipient organism.

Predicting the impact of HGT on the persistence landscape is difficult because quantification of HGT is experimentally challenging and thus estimates of HGT rates in natural microbiomes vary widely between $10^{-7}$ to $10^{-11}$ events per generation, per cell (Baur, Hanselmann et al. 1996, Jiang and Paul 1998, Sheppard, Beddis and Barraclough 2020, Moralez, Szenkiel et al. 2021). Transfer rates are best characterized for antibiotic resistance genes (ARGs), because they are highly selected for in the presence of antibiotics, the protein groups are well classified (Anthony, Burnham et al. 2021), there are well established tools and methods for identifying a positive transfer event (Ferreiro, Crook et al. 2018), and the level of surveillance in clinical and water treatment settings is high. Transfer of ARGs has been shown to be prevalent in gut microbiomes from humans (Anthony, Burnham et al. 2021) and mice (Moubareck, Bourgeois et al. 2003, Neil, Allard et al. 2020, Leon-Sampedro, DelaFuente et al. 2021), and has also been measured in environmental microbiomes such as sewage, biofilms, and soils (Forsberg, Patel et al. 2014, Klumper, Droumpali et al. 2014, Lecuyer, Bourassa et al. 2018, Li, Dechesne et al. 2018). Evidence of retention of even slightly beneficial genes following transfer suggests that a wider variety of transferable functions could contribute to a persistence landscape (van Dijk, Hogeweg et al. 2020).The impact of HGT on a persistence landscape is, however, modulated by the decreasing efficiency of transfer as the phylogenetic distance between the donor and recipient strains increases (Ravenhall, Skunca et al. 2015).

One mechanism of gene transduction to provide complementation is delivery of functions to the host by phage infection. Auxiliary viral genes (AVGs), defined as phage genes that have functions not directly related to the phage life cycle, are both prevalent and diverse within phage (Graham, Camargo et al. 2024). AVGs identified in phage can be functional, and the functions can complement traits of the infected bacteria (Lindell, Jaffe et al. 2005, Monier, Welsh et al. 2012, Wu, Smith et al. 2022, Wegner, Roitman et al. 2024), even expanding the host's innate ecological niche (Monier, Chambouvet et al. 2017). Because of the ubiquity of phage in microbial communities and the high frequency of virus-host interactions (estimated $10^{28}$ infections per day in seawater; Breitbart 2012), transduction of functional genes from phage to an introduced organism could be a significant factor impacting a persistence landscape (Suttle 1994, Fuhrman and Noble 1995, Suttle 2007).

Prediction of HGT potential to complement an absent function in an introduced organism could be undertaken by assessing the genetic potential of the target microbiome through functional annotation of the metagenome. These annotations could aid inference of the potential for genetic complementation, but a rate estimate would be very difficult to determine since it would depend on many interacting factors. These factors include uptake of exogenous DNA by an introduced organism (its competency), its proximity to bacteria with mechanisms for transfer, and the stochasticity of transfer interactions. The capacity for specific types of HGT from genomic or metagenomic sequences can be assessed through computational approaches, usually through identification of plasmid (Krawczyk, Lipinski and Dziembowski 2018, Redondo-Salvo, Fernandez-Lopez et al. 2020), conjugative elements (Wang, Liu et al. 2024), or phage (Roux, Enault et al. 2015, Camargo, Roux et al. 2023) sequences. However, these methods are likely limited in their accuracy and scope, especially when considering the number and diversity of plasmids and phages already known. Research to address this knowledge gap about HGT rates in natural communities would accelerate understanding of persistence landscapes and enhance bioengineering efforts toward genetic isolation of the target organism.

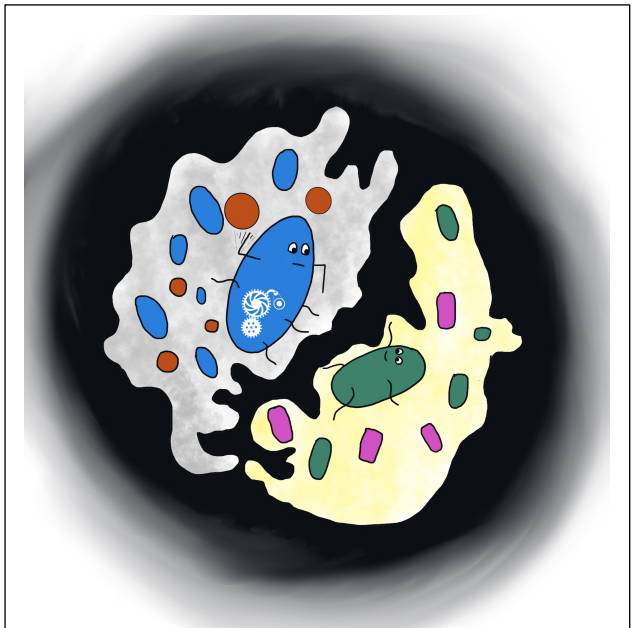

**Figure 6. Spatial distribution.** The distribution of community members across structurally isolated physical and chemical environments can impact the persistence landscape by providing or excluding critical nutrients or community interactions, synergistic or antagonistic.

## Environmental and Temporal Dynamics

The composition of microbiomes, and concordantly the relationships between members, are driven by the surrounding environment. The ability of a microbiome to complement absent functions will be shaped by the environment, with conditions that vary over time and space, changing the persistence landscape in both subtle and substantial ways depending on the type of perturbation or change. The relevant time and spatial scales to consider will vary based on the environment and may not be easy to determine. However, the following questions can guide identification of factors to constrain practical efforts to define a persistence landscape: Which conditions and/or environments are expected to be important for the persistence phenotype? How long must the introduced organism persist in the target community to serve its purpose? How are the boundaries of the target community defined and what scale of spatial measurements are needed to assess community and environmental parameters within those boundaries? What environmental perturbations might be expected to occur in this time and space? Defining a persistence landscape would require design of experiments to gather data with these core factors in mind.

Just as important to determining a persistence landscape is the intended purpose and time frame of the introduced function. If it is meant to be a short-lived function specific to a condition, the colonization side of the persistence landscape will be defined on a limited set of conditions and a short time scale. However, if the goal is long-term persistence of an introduced organism to provide a sustained or seasonal function, the persistence landscape will be defined on a wider set of possible conditions. In this case, considerations to prioritize include the anticipated deployment timescale for the introduced organism, the most critical conditions to assay for introduced function, and whether these conditions might vary over time. The interaction of the introduced organism with the most relevant conditions and time scale can thus be used to characterize the persistence landscape and use it to predict how the introduced function will be expressed under changing conditions.

Sudden shifts in the behavior of complex biological systems through biotic or abiotic perturbations have been studied extensively. Examples of perturbations include (i) a shift in environmental conditions beyond the normal range that results in mass turnover of community composition and its associated functions (Roy Chowdhury, Lee et al. 2019), or (ii) a phage bloom that alters community population dynamics (Zhao, Liang et al. 2025), potentially including the introduction of new functions through AVGs. These functional tipping points can be predicted for specific systems with sufficiently detailed models (Zhao, Liang et al. 2025), but this area of community modeling is not well developed. Perturbations can lead to large shifts in species abundance, community member functions, and community-level interactions, opening or closing environmental niche space that defines a persistence landscape.

Practical assessment of persistence landscape dynamics can be achieved through targeted experiments that sample the range of conditions expected, with appropriate temporal sampling. The landscape in this case would be the sampling aggregate of functional potential and complementation in the experimental set. The development and use of fabricated ecosystems for soils such as EcoFab (Gupta, Tian et al. 2024), the RhizoGrid (Handakumbura, Rivas Ubach and Battu 2021), and RhizoChip (Aufrecht, Khalid et al. 2022) that capture spatially-resolved data will allow the dynamics of these systems to be studied in situ. A powerful way to approach this problem is the use of model microbial consortia, referred to as synthetic communities (SynComs), that contain a small number of individual species that represent taxonomic and functional diversity present in complex communities in reproducible and experimentally tractable forms (McClure, Naylor et al. 2020, Fonseca-Garcia, Pettinga et al. 2024). SynComs that represent the most important members of a complex community can be subjected to a wide range of controlled experiments to understand the responses of native and amended communities in different environments. They enable the testing and characterization of a persistence landscape through experiments with an introduced organism.

## Spatial Distribution

The spatial structure of the environment and the arrangement of cells/populations within it constrain the interactions between community members, and thus the persistence landscape. The environmental matrix, from a host gut to a soil system, modulates the transport and accessibility of metabolites, phage, and cells through aqueous channels or other structural features that connect disparate regions of the environment. Thus, persistence is dependent on the spatial context – the proximity to the source of interactions which promote or prevent persistence. The complexity of spatial factors that influence interactions with potentially complementary functions makes it a challenging and active area of study, where emerging technologies and innovative approaches are being applied.

Community structure and interactions are frequently influenced by biofilms. Biofilms are intercellular polymeric matrices formed by production of extracellular polymeric substances (EPS), which allow aggregation of microbes into pore and channel structures. These structures bring microbes into close pockets and create environments that can drive formation of chemical gradients, but are generally protective of the bacteria inside, e.g. from antibiotics and other stresses (Bohning, Tarafder and Bharat 2024). Biofilms also constrain cell-cell interaction and communication via mechanisms such as quorum sensing and HGT (Lecuyer, Bourassa et al. 2018, Michielsen, Vercelli et al. 2024). For some communities, integration of an introduced organism into a biofilm might be advantageous or critical to persistence or successful delivery of its intended function. Improved understanding about how introduction of non-native organisms works at different stages of biofilm life cycle/development and/or methods of introduction is needed to better understand impact on the persistence landscape.

Spatial variation in human microbiomes has been elegantly characterized in a number of studies including a cartography of the human skin microbiome, which combined metagenomics and metabolomics (Bouslimani, Porto et al. 2015) to show that the skin exhibits diverse distributions of species. Similarly, a novel experimental platform, the RhizoGrid, was developed to map the bioenergy crop *Sorghum bicolor* root rhizosphere, showing that the distribution of bacterial taxa and metabolites vary across the root depth (Handakumbura, Rivas Ubach and Battu 2021). Both these examples highlight the difficulties of defining a persistence landscape for a microbiome, since the composition of the microbiome in the target environment is likely to be spatially heterogeneous. Focusing on a core microbiome (Shade and Handelsman 2012) from the target microbiome (e.g. root rhizosphere) is one approach to reduce variability across space. However, this can limit insight by ignoring the species and functions that are not shared but nevertheless play an important role in persistence. Efforts to profile the degree of compositional and/or functional variation over space could help address this hurdle. Fortunately, distance decay – the tendency for nearby locations to have similar characteristics or be influenced by each other (Wang, Zeng et al. 2024) – allows for implementing linear and non-linear forms of correlation and regression to identify microbes and metabolites that vary across, for instance, different rooting zones of the rhizosphere. Future efforts to identify causal relationships between components of the microbiome in different spatial regions could identify microbe-microbe, microbe-metabolite, and microbe-metabolite-host interactions that drive colonization dynamics of an introduced organism in natural settings which include a host like a plant or the human gut. Alongside experimental platforms like the RhizoGrid or Gut-on-a-chip platforms spatial enrichment and causal inference could be powerful tools for identifying how host and environmental perturbations affect the persistence landscape of complex microbiomes.

## Addressing Challenges to Defining Persistence Landscapes

The daunting complexity of composition layered by spatial and temporal interactions renders complete characterization of a persistence landscape for natural communities generally infeasible, even at the taxonomic composition level. Model systems, such as reduced-complexity SynComs (Duran, Thiergart et al. 2018, McClure, Naylor et al. 2020, van der Lelie, Oka et al. 2021, Fonseca-Garcia, Pettinga et al. 2024), offer one solution that has been applied to define a functional landscape (Sanchez-Gorostiaga, Bajic et al. 2019). Even though these simplified systems can be challenging due to the large number of potential interactions to investigate, SynComs can be used to examine species abundance profiles and correlations, which can then be used to identify keystone species and potential interacting partners. The community members can be more fully characterized individually to provide a more complete understanding of functional capabilities. As such, SynComs offer tools to study the behavior of communities and how they complement introduced organisms and can provide predictions of important molecular mechanisms that can be tested in more complex natural communities. Further, because they can be more deeply characterized than most natural communities, they are excellent test beds for evaluating the mechanisms that support the establishment and persistence control of introduced organisms.

Current limitations of metagenomic sequencing techniques and our knowledge of microbial physiology prevent full characterization of the taxonomic composition and functional potential of complex microbial communities such as the soil microbiome. We have previously published very deep sequencing of soil communities, and our results indicate that we have not yet approached determining all the species present (Nelson, Anderson et al. 2020). Species that are in very low abundance but that might thrive under specific conditions when assessed in a persistence landscape may confound efforts to fully understand their persistence potential. These issues mean that the ability to assemble complete or nearly complete genomes from metagenome reads is limited and it is not unusual that a large percentage of metagenome sequences cannot be incorporated into genomes (Thomas and Segata 2019). Additionally, the percentage of protein coding sequences that can be confidently assigned a function is between 30-40% even in relatively well-studied bacteria (Salzberg 2019, Lobb, Tremblay et al. 2020). Due to these challenges, there have been recent calls to the community to establish high quality domain-specific databases of microbially assembled genomes not often submitted to public databases (Anthony, Allison et al. 2024). This would increase data-reuse, promote Findable, Accessible, Interoperable, and Reusable (FAIR) metadata standards, and improve taxonomic and functional annotation for metagenomic "dark matter". Alongside this, novel machine learning methods for determining functional and metabolic potential trained on the extensive amount of available genome information is helping address these problems (Yu, Cui et al. 2023, Geller-McGrath, Konwar et al. 2024).

As described above, characterization of the persistence landscape requires understanding of community function, and the inter-organism relationships that arise from these functions. Genome sequencing provides information about functional potential but does not indicate directly which functions are active. Metatranscriptomics to measure gene expression is one step closer to the goal of assessing expressed function and can provide a good indication of the genes that are active under particular conditions. However, the expression of a well-annotated gene does not mean that the function it represents is active, because the relationship between mRNA levels and corresponding protein levels is complex and not always directly proportional (Zhang, Liu et al. 2016, Perl, Ushakov et al. 2017). Additionally, proteins can be post-translationally modified in many ways that modulate their function (Macek, Forchhammer et al. 2019). Metaproteomics (Maier, Lucio et al. 2017, Li, Wang et al. 2023) is a developing area that measures community protein abundance, providing a better estimate of function. However, the methodology is challenging, and as a result metaproteomics data tend to be sparse representations of the full community and do not capture all relevant functions (Pan, Wattiez and Gillan 2024). Multi-omic data integration, coupled with modeling, is a way of addressing these limitations by making use of the strengths of each data type (Deek, Ma et al. 2024). Improved methods for high-throughput phenotypic characterization, and better methods for real-time monitoring of molecular signatures are needed to better capture the complexity required for understanding the persistence landscape.

To construct models that are truly predictive of community capacity for supporting persistence of an introduced organism, we must improve modeling approaches. More work is needed to integrate ecological competition models, which are expected to be strongly influenced by the effects of functional complementation, with mechanistic metabolic models to understand how dynamics of this complementation impact the persistence landscape. Improved methods for community metabolic modeling that account for molecular and phenotypic measurements (McClure, Lee et al. 2020) may enable better estimation of a persistence landscape from the metabolic complementation perspective. Finally, working with data from synthetic model systems could allow advancement of these approaches, and provide frameworks for rapid iteration on model improvement, benchmarking, and data imputation.

## Practical Assessment and Application of a Rhizosphere Persistence Landscape

One important example of a target community for introduced organisms is the root rhizosphere, which is vitally important for plant growth (Afridi, Fakhar et al. 2022, Orozco-Mosqueda, Fadiji et al. 2022, Jansson, McClure and Egbert 2023). Understanding the persistence landscape of the root rhizosphere environment, including the metabolic complementation potential of the host plant, will enable enhanced design of engineered microbiomes. These enhancements include an increased ability to compete within native communities for host exudates and to deal with the important issues of containment and escape within an agricultural context.

Ethical application of engineered organisms to microbial communities will likely require the organism to be engineered to prevent spread beyond the target area to minimize unintended consequences. Thus, the persistence landscape can also be used to guide minimization of fitness in non-targeted environments. Mapping of the persistence landscape would inform strategies to increase colonization potential within the sorghum root environment, while preventing escape to the surrounding bulk soil microbiome. Within the scope of metabolic complementation, one goal could be to identify metabolites that are not produced by other microbial constituents but which are *solely* produced by the plant. For instance, in the bioenergy crop *Sorghum bicolor*, root exudates are largely comprised of a relatively small list of plant produced metabolites, several of which are, to the best of our knowledge, rarely if ever produced by other host plants or other microorganisms (Czarnota, Rimando and Weston 2003, Wang, Chai et al. 2021). Understanding which microbial genes enable the uptake and use of these compounds as carbon sources would provide a method for engineering persistence control into an introduced organism, for example to be dependent upon that carbon source for growth or to only express a function when that compound is present. Similar approaches could be envisioned in human microbiomes by using selective diet supplementation with rare carbon sources.

The root rhizosphere comprises many thousands of species interacting in a complex spatial environment, making a comprehensive description of a persistence landscape experimentally intractable. To extend the concept of the functional landscape to the persistence landscape we instead envision near-term experiments that focus on stable synthetic communities designed to represent aspects of the target community and an introduced, possibly engineered, organism. Following Sanchez-Gorostiaga, et al., the membership of the synthetic community would be varied and the persistence of an introduced organism measured as "persistence fitness". Enumerating all possible community compositions as unique combinations of species allows definition of the persistence landscape, identification of combinations of species permissive or intolerant of an introduced organism, and sub-groups of community members that fulfill different functional roles and niche specialties. Similar studies using synthetic communities designed to reflect aspects of non-target communities would be used to identify and characterize possible routes of escape for the introduced organism, providing a basis for refinement of control strategies. As experimental and modeling approaches advance, we anticipate more sophisticated opportunities to explore and define persistence landscapes.

Advances in automated experimentation using robotic platforms offer a way to approach the persistence landscape. Coupled with smart computational approaches to guide experiments, automated platforms can be used to characterize important areas of the landscape and lead to a predictive method for novel combinations. To our knowledge this has not been attempted with microbial communities, but a recent example of a similar framework involved a deep learning-powered, automated pipeline that learned the combination of amino acids supporting growth of a single organism (Dama, Kim et al. 2023). While the persistence landscape presents additional challenges including experimental consideration of members of the community that may be difficult or impossible to culture, and the much larger potential combinatorial space, a similar approach could be very useful for exploring persistence landscapes.

Realization of a persistence landscape would allow improved design of engineered organisms that could survive in a complex target community and carry out their intended function, while keeping them contained safely to that community. Experimental data would need to be coupled with improved models that more completely describe the persistence landscape and could be used to formulate new strategies for modulation of engineered functions. A more complete understanding of the persistence landscape would allow development of approaches to enhance crop yield, decrease dependence on chemical fertilizers, improve resilience against environmental changes, and secure crops against potential biosecurity threats.

**Acknowledgements**

This work was supported by the Department of Energy (DOE) Office of Biological and Environmental Research (BER) and is a contribution of the Scientific Focus Area "Persistence Control of Engineered Functions in Complex Soil Microbiomes". This work was authored in part by Oak Ridge National Laboratory, which is managed by UT-Battelle, LLC, for the U.S. Department of Energy under contract DE-AC05-00OR22725. The Pacific Northwest National Laboratory is operated for the DOE by Battelle Memorial Institute under Contract DE-AC05-76RL01830.

**References**

Afridi, M. S., A. Fakhar, A. Kumar, S. Ali, F. H. V. Medeiros, M. A. Muneer, H. Ali and M. Saleem (2022). "Harnessing microbial multitrophic interactions for rhizosphere microbiome engineering." Microbiol Res **265**: 127199.

Akhter, S. and J. H. Miller (2023). "BaPreS: a software tool for predicting bacteriocins using an optimal set of features." BMC Bioinformatics **24**(1): 313.

Ang, K. S., M. Lakshmanan, N. R. Lee and D. Y. Lee (2018). "Metabolic Modeling of Microbial Community Interactions for Health, Environmental and Biotechnological Applications." Curr Genomics **19**(8): 712-722.

Anthony, W. E., S. D. Allison, C. M. Broderick, L. Chavez Rodriguez, A. Clum, H. Cross, E. Eloe-Fadrosh, S. Evans, D. Fairbanks, R. Gallery, J. B. Gontijo, J. Jones, J. McDermott, J. Pett-Ridge, S. Record, J. L. M. Rodrigues, W. Rodriguez-Reillo, K. L. Shek, T. Takacs-Vesbach and J. L. Blanchard (2024). "From soil to sequence: filling the critical gap in genome-resolved metagenomics is essential to the future of soil microbial ecology." Environ Microbiome **19**(1): 56.

Anthony, W. E., C. D. Burnham, G. Dantas and J. H. Kwon (2021). "The Gut Microbiome as a Reservoir for Antimicrobial Resistance." J Infect Dis **223**(12 Suppl 2): S209-S213.

Aufrecht, J., M. Khalid, C. L. Walton, K. Tate, J. F. Cahill and S. T. Retterer (2022). "Hotspots of root-exuded amino acids are created within a rhizosphere-on-a-chip." Lab Chip **22**(5): 954-963.

Baumann, P. (2005). "Biology bacteriocyte-associated endosymbionts of plant sap-sucking insects." Annu Rev Microbiol **59**: 155-189.

Baur, B., K. Hanselmann, W. Schlimme and B. Jenni (1996). "Genetic transformation in freshwater: Escherichia coli is able to develop natural competence." Appl Environ Microbiol **62**(10): 3673-3678.

Bohning, J., A. K. Tarafder and T. A. M. Bharat (2024). "The role of filamentous matrix molecules in shaping the architecture and emergent properties of bacterial biofilms." Biochem J **481**(4): 245-263.

Bouslimani, A., C. Porto, C. M. Rath, M. Wang, Y. Guo, A. Gonzalez, D. Berg-Lyon, G. Ackermann, G. J. Moeller Christensen, T. Nakatsuji, L. Zhang, A. W. Borkowski, M. J. Meehan, K. Dorrestein, R. L. Gallo, N. Bandeira, R. Knight, T. Alexandrov and P. C. Dorrestein (2015). "Molecular cartography of the human skin surface in 3D." Proc Natl Acad Sci U S A **112**(17): E2120-2129.

Camargo, A. P., S. Roux, F. Schulz, M. Babinski, Y. Xu, B. Hu, P. S. G. Chain, S. Nayfach and N. C. Kyrpides (2023). "Identification of mobile genetic elements with geNomad." Nat Biotechnol.

Chan, C. T., J. W. Lee, D. E. Cameron, C. J. Bashor and J. J. Collins (2016). "'Deadman' and 'Passcode' microbial kill switches for bacterial containment." Nat Chem Biol **12**(2): 82-86.

Chassaing, B. and E. Cascales (2018). "Antibacterial Weapons: Targeted Destruction in the Microbiota." Trends Microbiol **26**(4): 329-338.

Coe, A., R. Braakman, S. J. Biller, A. Arellano, C. Bliem, N. N. Vo, K. von Emster, E. Thomas, M. DeMers, C. Steglich, J. Huisman and S. W. Chisholm (2024). "Emergence of metabolic coupling to the heterotroph Alteromonas promotes dark survival in Prochlorococcus." ISME Commun **4**(1): ycae131.

Costa, S. S., G. da Silva Moia, A. Silva, R. A. Barauna and A. A. de Oliveira Veras (2023). "BADASS: BActeriocin-Diversity ASsessment Software." BMC Bioinformatics **24**(1): 24.

Czarnota, M. A., A. M. Rimando and L. A. Weston (2003). "Evaluation of root exudates of seven sorghum accessions." Journal of Chemical Ecology 29(9): 2073-2083.

Dama, A. C., K. S. Kim, D. M. Leyva, A. P. Lunkes, N. S. Schmid, K. Jijakli and P. A. Jensen (2023). "BacterAI maps microbial metabolism without prior knowledge." Nat Microbiol 8(6): 1018-1025.

Deek, R. A., S. Ma, J. Lewis and H. Li (2024). "Statistical and computational methods for integrating microbiome, host genomics, and metabolomics data." Elife 13.

Diaz-Colunga, J., A. Skwara, K. Gowda, R. Diaz-Uriarte, M. Tikhonov, D. Bajic and A. Sanchez (2023). "Global epistasis on fitness landscapes." Philos Trans R Soc Lond B Biol Sci 378(1877): 20220053.

Douglas, G. M., V. J. Maffei, J. R. Zaneveld, S. N. Yurgel, J. R. Brown, C. M. Taylor, C. Huttenhower and M. G. I. Langille (2020). "PICRUSt2 for prediction of metagenome functions." Nat Biotechnol 38(6): 685-688.

Dragos, A., A. J. C. Andersen, C. N. Lozano-Andrade, P. J. Kempen, A. T. Kovacs and M. L. Strube (2021). "Phages carry interbacterial weapons encoded by biosynthetic gene clusters." Curr Biol 31(16): 3479-3489 e3475.

Duran, P., T. Thiergart, R. Garrido-Oter, M. Agler, E. Kemen, P. Schulze-Lefert and S. Hacquard (2018). "Microbial Interkingdom Interactions in Roots Promote Arabidopsis Survival." Cell 175(4): 973-983 e914.

Egbert, R. G., H. S. Rishi, B. A. Adler, D. M. McCormick, E. Toro, R. T. Gill and A. P. Arkin (2019). "A versatile platform strain for high-fidelity multiplex genome editing." Nucleic Acids Res 47(6): 3244-3256.

Elmore, J. R., G. N. Dexter, H. Baldino, J. D. Huenemann, R. Francis, G. L. t. Peabody, J. Martinez-Baird, L. A. Riley, T. Simmons, D. Coleman-Derr, A. M. Guss and R. G. Egbert (2023). "High-throughput genetic engineering of nonmodel and undomesticated bacteria via iterative site-specific genome integration." Sci Adv 9(10): eade1285.

Ershova, A. S., I. S. Rusinov, S. A. Spirin, A. S. Karyagina and A. V. Alexeevski (2015). "Role of Restriction-Modification Systems in Prokaryotic Evolution and Ecology." Biochemistry (Mosc) 80(10): 1373-1386.

Etherton, B. A., R. A. Choudhury, R. I. Alcala Briseno, R. A. Mouafo-Tchinda, A. I. Plex Sula, M. Choudhury, A. Adhikari, S. L. Lei, N. Kraisitudomsook, J. R. Buritica, V. A. Cerbaro, K. Ogero, C. M. Cox, S. P. Walsh, J. L. Andrade-Piedra, B. A. Omondi, I. Navarrete, M. A. McEwan and K. A. Garrett (2024). "Disaster Plant Pathology: Smart Solutions for Threats to Global Plant Health from Natural and Human-Driven Disasters." Phytopathology 114(5): 855-868.

Ferreiro, A., N. Crook, A. J. Gasparrini and G. Dantas (2018). "Multiscale Evolutionary Dynamics of Host-Associated Microbiomes." Cell 172(6): 1216-1227.

Fonseca-Garcia, C., D. Pettinga, A. Wilson, J. R. Elmore, R. McClure, J. Atim, J. Pedraza, R. Hutmacher, H. Turumtay, Y. Tian, A. Eudes, H. V. Scheller, R. G. Egbert and D. Coleman-Derr (2024). "Defined synthetic microbial communities colonize and benefit field-grown sorghum." ISME J 18(1).

Forsberg, K. J., S. Patel, M. K. Gibson, C. L. Lauber, R. Knight, N. Fierer and G. Dantas (2014). "Bacterial phylogeny structures soil resistomes across habitats." Nature 509(7502): 612-+.

Fuhrman, J. A. and R. T. Noble (1995). "Viruses and protists cause similar bacterial mortality in coastal seawater." Limnology and Oceanography 40(7): 1236-1242.

Garber, A. I., A. Garcia de la Filia Molina, I. M. Vea, A. J. Mongue, L. Ross and J. P. McCutcheon (2024). "Retention of an Endosymbiont for the Production of a Single Molecule." Genome Biol Evol 16(4).

Garber, A. I., M. Kupper, D. R. Laetsch, S. R. Weldon, M. S. Ladinsky, P. J. Bjorkman and J. P. McCutcheon (2021). "The Evolution of Interdependence in a Four-Way Mealybug Symbiosis." Genome Biol Evol **13**(8).

Geesink, P., J. Ter Horst and T. J. G. Ettema (2024). "More than the sum of its parts: uncovering emerging effects of microbial interactions in complex communities." FEMS Microbiol Ecol **100**(4).

Geller-McGrath, D., K. M. Konwar, V. P. Edgcomb, M. Pachiadaki, J. W. Roddy, T. J. Wheeler and J. E. McDermott (2024). "Predicting metabolic modules in incomplete bacterial genomes with MetaPathPredict." Elife **13**.

Ghequire, M. G. K. and R. De Mot (2015). "The Tailocin Tale: Peeling off Phage Tails." Trends Microbiol **23**(10): 587-590.

Gonze, D., K. Z. Coyte, L. Lahti and K. Faust (2018). "Microbial communities as dynamical systems." Curr Opin Microbiol **44**: 41-49.

Goyal, A., T. Wang, V. Dubinkina and S. Maslov (2021). "Ecology-guided prediction of cross-feeding interactions in the human gut microbiome." Nat Commun **12**(1): 1335.

Graham, E. B., A. P. Camargo, R. Wu, R. Y. Neches, M. Nolan, D. Paez-Espino, N. C. Kyrpides, J. K. Jansson, J. E. McDermott, K. S. Hofmockel and C. Soil Virosphere (2024). "A global atlas of soil viruses reveals unexplored biodiversity and potential biogeochemical impacts." Nat Microbiol **9**(7): 1873-1883.

Gupta, K., Y. Tian, A. Eudes, H. V. Scheller, A. K. Singh, P. D. Adams, P. F. Andeer and T. R. Northen (2024). "EcoFAB 3.0: a sterile system for studying sorghum that replicates previous field and greenhouse observations." Front Plant Sci **15**: 1440728.

Hammami, R., A. Zouhir, C. Le Lay, J. Ben Hamida and I. Fliss (2010). "BACTIBASE second release: a database and tool platform for bacteriocin characterization." BMC Microbiol **10**: 22.

Handakumbura, P. P., A. Rivas Ubach and A. K. Battu (2021). "Visualizing the Hidden Half: Plant-Microbe Interactions in the Rhizosphere." mSystems **6**(5): e0076521.

Haskett, T. L., A. Tkacz and P. S. Poole (2021). "Engineering rhizobacteria for sustainable agriculture." ISME J **15**(4): 949-964.

Henriques, S. F., D. B. Dhakan, L. Serra, A. P. Francisco, Z. Carvalho-Santos, C. Baltazar, A. P. Elias, M. Anjos, T. Zhang, O. D. K. Maddocks and C. Ribeiro (2020). "Metabolic cross-feeding in imbalanced diets allows gut microbes to improve reproduction and alter host behaviour." Nat Commun **11**(1): 4236.

Hu, Y., J. Cai, Y. Song, G. Li, Y. Gong, X. Jiang, X. Tang, K. Shao and G. Gao (2024). "Sediment DNA Records the Critical Transition of Bacterial Communities in the Arid Lake." Microb Ecol **87**(1): 68.

Huber, N., M. Andraud, E. L. Sassu, C. Prigge, V. Zoche-Golob, A. Kasbohrer, D. D'Angelantonio, A. Viltrop, J. Zmudzki, H. Jones, R. P. Smith, T. Tobias and E. Burow (2022). "What is a biosecurity measure? A definition proposal for animal production and linked processing operations." One Health **15**: 100433.

Hugerth, L. W. and A. F. Andersson (2017). "Analysing Microbial Community Composition through Amplicon Sequencing: From Sampling to Hypothesis Testing." Front Microbiol **8**: 1561.

Ikryannikova, L. N., L. K. Kurbatov, N. V. Gorokhovets and A. A. Zamyatnin, Jr. (2020). "Contact-Dependent Growth Inhibition in Bacteria: Do Not Get Too Close!" Int J Mol Sci **21**(21).

Isabella, V. M., B. N. Ha, M. J. Castillo, D. J. Lubkowicz, S. E. Rowe, Y. A. Millet, C. L. Anderson, N. Li, A. B. Fisher, K. A. West, P. J. Reeder, M. M. Momin, C. G. Bergeron, S. E. Guilmain, P. F. Miller, C. B. Kurtz and D. Falb (2018). "Development of a synthetic live bacterial therapeutic for the human metabolic disease phenylketonuria." Nature Biotechnology **36**(9): 857-864.

Jaderlund, L., M. Hellman, I. Sundh, M. J. Bailey and J. K. Jansson (2008). "Use of a novel nonantibiotic triple marker gene cassette to monitor high survival of Pseudomonas fluorescens SBW25 on winter wheat in the field." FEMS Microbiol Ecol **63**(2): 156-168.

Jansson, J. K., R. McClure and R. G. Egbert (2023). "Soil microbiome engineering for sustainability in a changing environment." Nat Biotechnol **41**(12): 1716-1728.

Jiang, S. C. and J. H. Paul (1998). "Gene transfer by transduction in the marine environment." Appl Environ Microbiol **64**(8): 2780-2787.

Joseph, C., H. Zafeiropoulos, K. Bernaerts and K. Faust (2024). "Predicting microbial interactions with approaches based on flux balance analysis: an evaluation." BMC Bioinformatics **25**(1): 36.

Kauffman, S. and S. Levin (1987). "Towards a general theory of adaptive walks on rugged landscapes." J Theor Biol **128**(1): 11-45.

Klumper, U., A. Droumpali, A. Dechesne and B. F. Smets (2014). "Novel assay to measure the plasmid mobilizing potential of mixed microbial communities." Front Microbiol **5**: 730.

Krawczyk, P. S., L. Lipinski and A. Dziembowski (2018). "PlasFlow: predicting plasmid sequences in metagenomic data using genome signatures." Nucleic Acids Res **46**(6): e35.

Lange, E., L. Kranert, J. Kruger, D. Benndorf and R. Heyer (2024). "Microbiome modeling: a beginner's guide." Front Microbiol **15**: 1368377.

Laub, M. T. and A. Typas (2024). "Principles of bacterial innate immunity against viruses." Curr Opin Immunol **89**: 102445.

Lecuyer, F., J. S. Bourassa, M. Gelinas, V. Charron-Lamoureux, V. Burrus and P. B. Beauregard (2018). "Biofilm Formation Drives Transfer of the Conjugative Element ICEBs1 in Bacillus subtilis." mSphere **3**(5).

Lee, J. Y., S. Haruta, S. Kato, H. C. Bernstein, S. R. Lindemann, D. Y. Lee, J. K. Fredrickson and H. S. Song (2019). "Prediction of Neighbor-Dependent Microbial Interactions From Limited Population Data." Front Microbiol **10**: 3049.

Leon-Sampedro, R., J. DelaFuente, C. Diaz-Agero, T. Crellen, P. Musicha, J. Rodriguez-Beltran, C. de la Vega, M. Hernandez-Garcia, R. G. W. S. Group, N. Lopez-Fresnena, P. Ruiz-Garbajosa, R. Canton, B. S. Cooper and A. San Millan (2021). "Pervasive transmission of a carbapenem resistance plasmid in the gut microbiota of hospitalized patients." Nat Microbiol **6**(5): 606-616.

Leventhal, D. S., A. Sokolovska, N. Li, C. Plescia, S. A. Kolodziej, C. W. Gallant, R. Christmas, J. R. Gao, M. J. James, A. Abin-Fuentes, M. Momin, C. Bergeron, A. Fisher, P. F. Miller, K. A. West and J. M. Lora (2020). "Immunotherapy with engineered bacteria by targeting the STING pathway for anti-tumor immunity." Nature Communications **11**(1): 2739.

Li, L., T. Wang, Z. Ning, X. Zhang, J. Butcher, J. M. Serrana, C. M. A. Simopoulos, J. Mayne, A. Stintzi, D. R. Mack, Y. Y. Liu and D. Figeys (2023). "Revealing proteome-level functional redundancy in the human gut microbiome using ultra-deep metaproteomics." Nat Commun **14**(1): 3428.

Li, L. G., A. Dechesne, Z. M. He, J. S. Madsen, J. Nesme, S. J. Sorensen and B. F. Smets (2018). "Estimating the Transfer Range of Plasmids Encoding Antimicrobial Resistance in a Wastewater Treatment Plant Microbial Community." Environmental Science & Technology Letters **5**(5): 260-265.

Lindell, D., J. D. Jaffe, Z. I. Johnson, G. M. Church and S. W. Chisholm (2005). "Photosynthesis genes in marine viruses yield proteins during host infection." Nature **438**(7064): 86-89.

Lobb, B., B. J. Tremblay, G. Moreno-Hagelsieb and A. C. Doxey (2020). "An assessment of genome annotation coverage across the bacterial tree of life." Microb Genom **6**(3).

Macek, B., K. Forchhammer, J. Hardouin, E. Weber-Ban, C. Grangeasse and I. Mijakovic (2019). "Protein post-translational modifications in bacteria." Nat Rev Microbiol 17(11): 651-664.

Maier, T. V., M. Lucio, L. H. Lee, N. C. VerBerkmoes, C. J. Brislawn, J. Bernhardt, R. Lamendella, J. E. McDermott, N. Bergeron, S. S. Heinzmann, J. T. Morton, A. Gonzalez, G. Ackermann, R. Knight, K. Riedel, R. M. Krauss, P. Schmitt-Kopplin and J. K. Jansson (2017). "Impact of Dietary Resistant Starch on the Human Gut Microbiome, Metaproteome, and Metabolome." mBio 8(5).

Marken, J., M. Maxon and R. Murray (2024). Policy Recommendations for the Regulation of Engineered Microbes for Environmental Release, The Linde Center for Science, Society, and Policy, Caltech.

Martiny, A. C., K. Treseder and G. Pusch (2013). "Phylogenetic conservatism of functional traits in microorganisms." ISME J 7(4): 830-838.

Martiny, J. B., S. E. Jones, J. T. Lennon and A. C. Martiny (2015). "Microbiomes in light of traits: A phylogenetic perspective." Science 350(6261): aac9323.

Mawarda, P. C., S. L. Lakke, J. Dirk van Elsas and J. F. Salles (2022). "Temporal dynamics of the soil bacterial community following Bacillus invasion." iScience 25(5): 104185.

McClure, R., Y. Farris, R. Danczak, W. Nelson, H. S. Song, A. Kessell, J. Y. Lee, S. Couvillion, C. Henry, J. K. Jansson and K. S. Hofmockel (2022). "Interaction Networks Are Driven by Community-Responsive Phenotypes in a Chitin-Degrading Consortium of Soil Microbes." mSystems 7(5): e0037222.

McClure, R., D. Naylor, Y. Farris, M. Davison, S. J. Fansler, K. S. Hofmockel and J. K. Jansson (2020). "Development and Analysis of a Stable, Reduced Complexity Model Soil Microbiome." Front Microbiol 11: 1987.

McClure, R. S., J. Y. Lee, T. R. Chowdhury, E. M. Bottos, R. A. White, 3rd, Y. M. Kim, C. D. Nicora, T. O. Metz, K. S. Hofmockel, J. K. Jansson and H. S. Song (2020). "Integrated network modeling approach defines key metabolic responses of soil microbiomes to perturbations." Sci Rep 10(1): 10882.

McClure, R. S., C. C. Overall, E. A. Hill, H. S. Song, M. Charania, H. C. Bernstein, J. E. McDermott and A. S. Beliaev (2018). "Species-specific transcriptomic network inference of interspecies interactions." ISME J 12(8): 2011-2023.

McCutcheon, J. P., A. I. Garber, N. Spencer and J. M. Warren (2024). "How do bacterial endosymbionts work with so few genes?" PLoS Biol 22(4): e3002577.

McCutcheon, J. P. and Y. Lekberg (2019). "Symbiosis: Fungi as Shrewd Trade Negotiators." Curr Biol 29(12): R570-R572.

McCutcheon, J. P. and N. A. Moran (2011). "Extreme genome reduction in symbiotic bacteria." Nat Rev Microbiol 10(1): 13-26.

Michielsen, S., G. T. Vercelli, O. X. Cordero and H. Bachmann (2024). "Spatially structured microbial consortia and their role in food fermentations." Curr Opin Biotechnol 87: 103102.

Miller, D. P., Z. R. Fitzsimonds and R. J. Lamont (2019). "Metabolic Signaling and Spatial Interactions in the Oral Polymicrobial Community." J Dent Res 98(12): 1308-1314.

Mirzaei, S. and M. Tefagh (2024). "GEM-based computational modeling for exploring metabolic interactions in a microbial community." PLoS Comput Biol 20(6): e1012233.

Mondal, R. K., D. Sen, A. Arya and S. K. Samanta (2023). "Developing anti-microbial peptide database version 1 to provide comprehensive and exhaustive resource of manually curated AMPs." Sci Rep 13(1): 17843.

Monier, A., A. Chambouvet, D. S. Milner, V. Attah, R. Terrado, C. Lovejoy, H. Moreau, A. E. Santoro, E. Derelle and T. A. Richards (2017). "Host-derived viral transporter protein for nitrogen uptake in infected marine phytoplankton." Proc Natl Acad Sci U S A 114(36): E7489-E7498.

Monier, A., R. M. Welsh, C. Gentemann, G. Weinstock, E. Sodergren, E. V. Armbrust, J. A. Eisen and A. Z. Worden (2012). "Phosphate transporters in marine phytoplankton and their viruses: cross-domain commonalities in viral-host gene exchanges." Environ Microbiol **14**(1): 162-176.

Moralez, J., K. Szenkiel, K. Hamilton, A. Pruden and A. J. Lopatkin (2021). "Quantitative analysis of horizontal gene transfer in complex systems." Curr Opin Microbiol **62**: 103-109.

Mori, M., M. Ponce-de-Leon, J. Pereto and F. Montero (2016). "Metabolic Complementation in Bacterial Communities: Necessary Conditions and Optimality." Front Microbiol **7**: 1553.

Moubareck, C., N. Bourgeois, P. Courvalin and F. Doucet-Populaire (2003). "Multiple antibiotic resistance gene transfer from animal to human enterococci in the digestive tract of gnotobiotic mice." Antimicrob Agents Chemother **47**(9): 2993-2996.

Musat, N., F. Musat, P. K. Weber and J. Pett-Ridge (2016). "Tracking microbial interactions with NanoSIMS." Curr Opin Biotechnol **41**: 114-121.

National Academies of Sciences, Engineering, and Medicine (2019). Science breakthroughs to advance food and agricultural research by 2030, National Academies Press.

Neil, K., N. Allard, F. Grenier, V. Burrus and S. Rodrigue (2020). "Highly efficient gene transfer in the mouse gut microbiota is enabled by the IncI(2) conjugative plasmid TP114." Commun Biol **3**(1): 523.

Nelson, W. C., L. N. Anderson, R. Wu, J. E. McDermott, S. L. Bell, A. Jumpponen, S. J. Fansler, K. J. Tyrrell, Y. Farris, K. S. Hofmockel and J. K. Jansson (2020). "Terabase Metagenome Sequencing of Grassland Soil Microbiomes." Microbiol Resour Announc **9**(32).

Newman, N. K., M. S. Macovsky, R. R. Rodrigues, A. M. Bruce, J. W. Pederson, J. Padiadpu, J. Shan, J. Williams, S. S. Patil, A. K. Dzutsev, N. Shulzhenko, G. Trinchieri, K. Brown and A. Morgun (2024). "Transkingdom Network Analysis (TkNA): a systems framework for inferring causal factors underlying host-microbiota and other multi-omic interactions." Nat Protoc **19**(6): 1750-1778.

Nikaido, H. and J. M. Pages (2012). "Broad-specificity efflux pumps and their role in multidrug resistance of Gram-negative bacteria." FEMS Microbiol Rev **36**(2): 340-363.

Nuccio, E. E., S. J. Blazewicz, M. Lafler, A. N. Campbell, A. Kakouridis, J. A. Kimbrel, J. Wollard, D. Vyshenska, R. Riley, A. Tomatsu, R. Hestrin, R. R. Malmstrom, M. Firestone and J. Pett-Ridge (2022). "HT-SIP: a semi-automated stable isotope probing pipeline identifies cross-kingdom interactions in the hyphosphere of arbuscular mycorrhizal fungi." Microbiome **10**(1): 199.

Orozco-Mosqueda, M. D. C., A. E. Fadiji, O. O. Babalola, B. R. Glick and G. Santoyo (2022). "Rhizobiome engineering: Unveiling complex rhizosphere interactions to enhance plant growth and health." Microbiol Res **263**: 127137.

Pan, H., R. Wattiez and D. Gillan (2024). "Soil Metaproteomics for Microbial Community Profiling: Methodologies and Challenges." Curr Microbiol **81**(8): 257.

Pepe-Ranney, C., A. N. Campbell, C. N. Koechli, S. Berthrong and D. H. Buckley (2016). "Unearthing the Ecology of Soil Microorganisms Using a High Resolution DNA-SIP Approach to Explore Cellulose and Xylose Metabolism in Soil." Front Microbiol **7**: 703.

Perl, K., K. Ushakov, Y. Pozniak, O. Yizhar-Barnea, Y. Bhonker, S. Shivatzki, T. Geiger, K. B. Avraham and R. Shamir (2017). "Reduced changes in protein compared to mRNA levels across non-proliferating tissues." BMC Genomics **18**(1): 305.

Pivot Bio (2019). Pivot Bio Proven: 2019 Performance Report.

Qiu, Z., E. Egidi, H. Liu, S. Kaur and B. K. Singh (2019). "New frontiers in agriculture productivity: Optimised microbial inoculants and in situ microbiome engineering." Biotechnol Adv **37**(6): 107371.

Ravenhall, M., N. Skunca, F. Lassalle and C. Dessimoz (2015). "Inferring horizontal gene transfer." PLoS Comput Biol **11**(5): e1004095.

Reardon, S. (2018). "Genetically modified bacteria enlisted in fight against disease." Nature **558**(7711): 497-498.

Redondo-Salvo, S., R. Fernandez-Lopez, R. Ruiz, L. Vielva, M. de Toro, E. P. C. Rocha, M. P. Garcillan-Barcia and F. de la Cruz (2020). "Pathways for horizontal gene transfer in bacteria revealed by a global map of their plasmids." Nat Commun **11**(1): 3602.

Rives, A., J. Meier, T. Sercu, S. Goyal, Z. Lin, J. Liu, D. Guo, M. Ott, C. L. Zitnick, J. Ma and R. Fergus (2021). "Biological structure and function emerge from scaling unsupervised learning to 250 million protein sequences." Proc Natl Acad Sci U S A **118**(15).

Rottinghaus, A. G., A. Ferreiro, S. R. S. Fishbein, G. Dantas and T. S. Moon (2022). "Genetically stable CRISPR-based kill switches for engineered microbes." Nat Commun **13**(1): 672.

Roux, S., F. Enault, B. L. Hurwitz and M. B. Sullivan (2015). "VirSorter: mining viral signal from microbial genomic data." PeerJ **3**: e985.

Roy Chowdhury, T., J. Y. Lee, E. M. Bottos, C. J. Brislawn, R. A. White, 3rd, L. M. Bramer, J. Brown, J. D. Zucker, Y. M. Kim, A. Jumpponen, C. W. Rice, S. J. Fansler, T. O. Metz, L. A. McCue, S. J. Callister, H. S. Song and J. K. Jansson (2019). "Metaphenomic Responses of a Native Prairie Soil Microbiome to Moisture Perturbations." mSystems **4**(4).

Ruiz-Perez, D., I. Gimon, M. Sazal, K. Mathee and G. Narasimhan (2024). "Unfolding and de-confounding: biologically meaningful causal inference from longitudinal multi-omic networks using METALICA." mSystems **9**(10): e0130323.

Salzberg, S. L. (2019). "Next-generation genome annotation: we still struggle to get it right." Genome Biol **20**(1): 92.

Sanchez, A., D. Bajic, J. Diaz-Colunga, A. Skwara, J. C. C. Vila and S. Kuehn (2023). "The community-function landscape of microbial consortia." Cell Syst **14**(2): 122-134.

Sanchez-Gorostiaga, A., D. Bajic, M. L. Osborne, J. F. Poyatos and A. Sanchez (2019). "High-order interactions distort the functional landscape of microbial consortia." PLoS Biol **17**(12): e3000550.

Shade, A. and J. Handelsman (2012). "Beyond the Venn diagram: the hunt for a core microbiome." Environ Microbiol **14**(1): 4-12.

Sheppard, R. J., A. E. Beddis and T. G. Barraclough (2020). "The role of hosts, plasmids and environment in determining plasmid transfer rates: A meta-analysis." Plasmid **108**: 102489.

Sloan, D. B. and N. A. Moran (2012). "Genome reduction and co-evolution between the primary and secondary bacterial symbionts of psyllids." Mol Biol Evol **29**(12): 3781-3792.

Spigaglia, P. (2024). "Clostridioides difficile and Gut Microbiota: From Colonization to Infection and Treatment." Pathogens **13**(8).

Stein, R. R., V. Bucci, N. C. Toussaint, C. G. Buffie, G. Ratsch, E. G. Pamer, C. Sander and J. B. Xavier (2013). "Ecological modeling from time-series inference: insight into dynamics and stability of intestinal microbiota." PLoS Comput Biol **9**(12): e1003388.

Steinway, S. N., M. B. Biggs, T. P. Loughran, Jr., J. A. Papin and R. Albert (2015). "Inference of Network Dynamics and Metabolic Interactions in the Gut Microbiome." PLoS Comput Biol **11**(5): e1004338.

Suttle, C. A. (1994). "The significance of viruses to mortality in aquatic microbial communities." Microb Ecol **28**(2): 237-243.

Suttle, C. A. (2007). "Marine viruses--major players in the global ecosystem." Nat Rev Microbiol **5**(10): 801-812.

Temme, K., A. Tamsir, S. Bloch, R. Clark, E. Tung, K. Hammill, D. Higgins and A. Davis-Richardson (2019). Methods and Compositions for Improving Plant Traits.

Thomas, A. M. and N. Segata (2019). "Multiple levels of the unknown in microbiome research." BMC Biol **17**(1): 48.

Tokuda, M. and M. Shintani (2024). "Microbial evolution through horizontal gene transfer by mobile genetic elements." Microb Biotechnol **17**(1): e14408.

van der Lelie, D., A. Oka, S. Taghavi, J. Umeno, T. J. Fan, K. E. Merrell, S. D. Watson, L. Ouellette, B. Liu, M. Awoniyi, Y. Lai, L. Chi, K. Lu, C. S. Henry and R. B. Sartor (2021). "Rationally designed bacterial consortia to treat chronic immune-mediated colitis and restore intestinal homeostasis." Nat Commun **12**(1): 3105.

van Dijk, B., P. Hogeweg, H. M. Doekes and N. Takeuchi (2020). "Slightly beneficial genes are retained by bacteria evolving DNA uptake despite selfish elements." Elife **9**.

Van Leuven, J. T., M. Mao, D. D. Xing, G. M. Bennett and J. P. McCutcheon (2019). "Cicada Endosymbionts Have tRNAs That Are Correctly Processed Despite Having Genomes That Do Not Encode All of the tRNA Processing Machinery." mBio **10**(3).

Van Leuven, J. T., R. C. Meister, C. Simon and J. P. McCutcheon (2014). "Sympatric speciation in a bacterial endosymbiont results in two genomes with the functionality of one." Cell **158**(6): 1270-1280.

Vo, P. L. H., C. Ronda, S. E. Klompe, E. E. Chen, C. Acree, H. H. Wang and S. H. Sternberg (2021). "CRISPR RNA-guided integrases for high-efficiency, multiplexed bacterial genome engineering." Nat Biotechnol **39**(4): 480-489.

Wang, M., G. Liu, M. Liu, C. Tai, Z. Deng, J. Song and H. Y. Ou (2024). "ICEberg 3.0: functional categorization and analysis of the integrative and conjugative elements in bacteria." Nucleic Acids Res **52**(D1): D732-D737.

Wang, P., Y. N. Chai, R. Roston, F. E. Dayan and D. P. Schachtman (2021). "The Root Exudate Sorgoleone Shapes Bacterial Communities and Delays Network Formation." Msystems **6**(2).

Wang, X., J. Zeng, F. Chen, Z. Wang, H. Liu, Q. Zhang, W. Liu, W. Wang, Y. Guo, Y. Niu, L. Yuan, C. Ren, G. Yang, Z. Zhong and X. Han (2024). "Aridity shapes distinct biogeographic and assembly patterns of forest soil bacterial and fungal communities at the regional scale." Sci Total Environ **948**: 174812.

Wegner, H., S. Roitman, A. Kupczok, V. Braun, J. N. Woodhouse, H. P. Grossart, S. Zehner, O. Beja and N. Frankenberg-Dinkel (2024). "Identification of Shemin pathway genes for tetrapyrrole biosynthesis in bacteriophage sequences from aquatic environments." Nat Commun **15**(1): 8783.

Wright, S. (1932). "The roles of mutation, inbreeding, crossbreeding, and selection in evolution." Proceedings of the Sixth International Congress on Genetics **1**(8): 355-366.

Wu, L., H. J. Wu, J. Qiao, X. Gao and R. Borriss (2015). "Novel Routes for Improving Biocontrol Activity of Bacillus Based Bioinoculants." Front Microbiol **6**: 1395.

Wu, R., C. A. Smith, G. W. Buchko, I. K. Blaby, D. Paez-Espino, N. C. Kyrpides, Y. Yoshikuni, J. E. McDermott, K. S. Hofmockel, J. R. Cort and J. K. Jansson (2022). "Structural characterization of a soil viral auxiliary metabolic gene product - a functional chitosanase." Nat Commun **13**(1): 5485.

Wu, R., M. Velickovic and K. E. Burnum-Johnson (2024). "From single cell to spatial multi-omics: unveiling molecular mechanisms in dynamic and heterogeneous systems." Curr Opin Biotechnol **89**: 103174.

Wu, Y., X. Pang, Y. Wu, X. Liu and X. Zhang (2022). "Enterocins: Classification, Synthesis, Antibacterial Mechanisms and Food Applications." Molecules **27**(7).

Yu, T., H. Cui, J. C. Li, Y. Luo, G. Jiang and H. Zhao (2023). "Enzyme function prediction using contrastive learning." Science **379**(6639): 1358-1363.

Zelezniak, A., S. Andrejev, O. Ponomarova, D. R. Mende, P. Bork and K. R. Patil (2015). "Metabolic dependencies drive species co-occurrence in diverse microbial communities." Proc Natl Acad Sci U S A **112**(20): 6449-6454.

Zhang, H., T. Liu, Z. Zhang, S. H. Payne, B. Zhang, J. E. McDermott, J. Y. Zhou, V. A. Petyuk, L. Chen, D. Ray, S. Sun, F. Yang, L. Chen, J. Wang, P. Shah, S. W. Cha, P. Aiyetan, S. Woo, Y. Tian, M. A. Gritsenko, T. R. Clauss, C. Choi, M. E. Monroe, S. Thomas, S. Nie, C. Wu, R. J. Moore, K. H. Yu, D. L. Tabb, D. Fenyo, V. Bafna, Y. Wang, H. Rodriguez, E. S. Boja, T. Hiltke, R. C. Rivers, L. Sokoll, H. Zhu, I. M. Shih, L. Cope, A. Pandey, B. Zhang, M. P. Snyder, D. A. Levine, R. D. Smith, D. W. Chan, K. D. Rodland and C. Investigators (2016). "Integrated Proteogenomic Characterization of Human High-Grade Serous Ovarian Cancer." Cell **166**(3): 755-765.

Zhao, X., X. Liang, Z. Zhu, Z. Yuan, S. Yu, Y. Liu, J. Wang, K. Mason-Jones, Y. Kuzyakov, J. Chen, T. Ge and S. Wang (2025). "Phages Affect Soil Dissolved Organic Matter Mineralization by Shaping Bacterial Communities." Environ Sci Technol **59**(4): 2070-2081.

Zomorrodi, A. R. and C. D. Maranas (2012). "OptCom: a multi-level optimization framework for the metabolic modeling and analysis of microbial communities." PLoS Comput Biol **8**(2): e1002363.

Zuniga, A., F. Fuente, F. Federici, C. Lionne, J. Bonnet, V. de Lorenzo and B. Gonzalez (2018). "An Engineered Device for Indoleacetic Acid Production under Quorum Sensing Signals Enables Cupriavidus pinatubonensis JMP134 To Stimulate Plant Growth." ACS Synth Biol **7**(6): 1519-1527.